\documentclass[twocolumn]{aastex631}

\shorttitle{Cannibals in PARADISE}
\shortauthors{Niemela et al.}

\begin{document}

\title{Cannibals in PARADISE: The effect of merging interplanetary shocks on solar energetic particle events}

\author[0000-0002-3746-9246]{Antonio Niemela}
\affiliation{Centre for mathematical Plasma Astrophysics, Dept.\ of Mathematics, KU Leuven, B-3001 Leuven, Belgium.}
\affiliation{Solar-Terrestrial Centre of Excellence—SIDC, Royal Observatory of Belgium, B-1180 Brussels, Belgium.}

\author[0000-0001-6344-6956]{Nicolas Wijsen}
\affiliation{Centre for mathematical Plasma Astrophysics, Dept.\ of Mathematics, KU Leuven, B-3001 Leuven, Belgium.}

\author[0000-0003-1539-7832]{Angels Aran}
\affiliation{Dep. F\'{\i}sica Qu\`antica i Astrof\'{\i}sica. Institut de Ci\`encies del Cosmos (ICCUB), Universitat de Barcelona, Martí i Franqu\`es 1, E-08028 Barcelona, Spain.}
\affiliation{Institut d’Estudis Espacials de Catalunya (IEEC), Barcelona, Spain.}

\author[0000-0002-6097-374X]{Luciano Rodriguez}
\affiliation{Solar-Terrestrial Centre of Excellence—SIDC, Royal Observatory of Belgium, B-1180 Brussels, Belgium.}

\author[0000-0003-1169-3722]{Jasmina Magdalenic}
\affiliation{Centre for mathematical Plasma Astrophysics, Dept.\ of Mathematics, KU Leuven, B-3001 Leuven, Belgium.}
\affiliation{Solar-Terrestrial Centre of Excellence—SIDC, Royal Observatory of Belgium, B-1180 Brussels, Belgium.}

\author[0000-0002-1743-0651]{Stefaan Poedts}
\affiliation{Centre for mathematical Plasma Astrophysics, Dept.\ of Mathematics, KU Leuven, B-3001 Leuven, Belgium.}
\affiliation{Institute of Physics, University of Maria Curie-Skłodowska, ul.\ Radziszewskiego 10, PL-20-031 Lublin, Poland.}

\begin{abstract}

    Gradual solar energetic particle (SEP) events are associated with shocks driven by coronal mass ejections (CMEs). The merging of two CMEs (so-called Cannibalistic CMEs) and the interaction of their associated shocks, has been linked to some of the most powerful solar storms ever recorded. Multiple studies have focused on the observational aspects of these SEP events, yet only a handful have focused on modeling similar CME-CME interactions in the heliosphere using advanced magnetohydrodynamic (MHD) models. This work presents, to our knowledge, the first modeling results of a fully time-dependent 3D simulation that captures both the interaction of two CMEs and its effect on the acceleration and transport of SEPs. This is achieved by using an MHD model for the solar wind and CME propagation together with an integrated SEP model. We perform different simulations and compare the behavior of the energetic protons in three different solar wind environments, where a combination of two SEP-accelerating CMEs are modeled. We find that particle acceleration is significantly affected by the presence of both CMEs in the simulation. Initially, less efficient acceleration results in lower energy particles. However, as the CMEs converge and their shocks eventually merge, particle acceleration is significantly enhanced through multiple acceleration processes between CME-driven shocks, resulting in higher particle intensities and energy levels.
\end{abstract}


\section{Introduction} \label{sec:intro}

Solar energetic particles (SEPs) are considered to be among the most hazardous space weather phenomena in the heliosphere \citep{2015Schrijver}. Gradual SEP events may last several days and are believed to be the result of continuous particle acceleration at shock waves through, for example, diffusive shock acceleration \citep[DSA;][]{1977Axford,1978Bell,1978Blandford}. These SEP events accelerated by coronal mass ejection (CMEs) driven shocks \citep[see e.g,][and references therein]{1999Reames, 2016Desai} are the focus of this work.

\cite{2004Gopalswamy} analyzed large\footnote{A large event is defined as events showing intensity~$\geq$~10~proton~cm$^{-2}$~s$^{-1}$~sr$^{-1}$ in the $>$10~MeV channel of the GOES/SEM instrument.} SEP events of the solar cycle 23 and found that higher intensity SEP events were reached in cases where CMEs were preceded by other CMEs from the same solar source region. Using 3D analysis techniques with events in solar cycle 24, \citet{2020Zhuang} found that in interacting events where the fast CME is close to the slow preceding CME, the latter will precondition the ambient wind for the fast CME, resulting in more intense SEP events. Moreover, using a 1D analytical model for converging shocks in the corona, \citet{2019Wang} found that such a shock interaction may explain the intense ground-level enhancement (GLE) event observed on 2012 May 17.

However, due to the three-dimensional nature of the CME-CME interactions, the CME kinematics can greatly vary depending on the CME part that is considered. Namely, the CME-driven shocks can have different kinematic properties than the core of the flux-rope \citep{2014Temmer}. Moreover, these characteristics can significantly change after a CME collision \citep{2017Feng}. In such complex cases, modeling is a very useful tool for understanding the physics behind CME-CME interaction. Many studies have focused on modeling CME-CME interactions, and a comprehensive review is presented by \citet{2017Lugaz}.

Since the review by \citet{2017Lugaz}, new and improved CME models have been used to study CME-CME interactions \citep[e.g.,][]{Scolini2020,2021Telloni,2023Pal}. However, few studies have investigated energetic particle acceleration and transport in complex situations involving CME-CME interactions within the inner heliosphere. In \citet{2022Palmerio}, a multi-spacecraft study of a real case was presented, with a Research-to-Operations-to-Research (R2O2R) approach. The authors used Enlil \citep{2003ODSTRCIL} and Solar Energetic Particle Model \citep[SEPMOD, ][]{2007Luhmann,2010Luhmann,2017Luhmann} to model SEP intensities at different points in the heliosphere. However, since SEPMOD is primarily aimed at efficient SEP forecasting, it does not include processes like particle scattering and acceleration. 

Another recent case study was presented in \citet{2023Niemela}. The authors used the EUropean Heliospheric FORecasitng Information Asset \citep[EUHFORIA;][]{Pomoell2018} and PArticle Radiation Asset Directed at Interplanetary Space Exploration \citep[PARADISE;][]{Wijsen2019(1),2020WijsenPhDThesis} to propagate energetic particles (injected at the shock of a fast and wide CME) through a complex ambient solar wind with multiple preceding CMEs. They explored the effects of using different CME models on particle transport by analyzing intensity-time profiles at multiple spacecraft. Apart from PARADISE, other models that have coupled a 3D MHD solar wind code with a particle transport code include, for example, Multiple-Field-Line-Advection Model for Particle Acceleration \citep[M-FLAMPA,][]{2018Borovikov}, the improved Particle Acceleration and Transport in the Heliosphere \citep[iPATH,][]{2022Ding_b}, and the Energetic Particle Radiation Environment Module \citep[EPREM,][]{2021Young}.  However, to the best of our knowledge, these models have not yet been utilized to study interacting CMEs and their effect on the associated SEP event.

Our study focuses on the large-scale acceleration processes in the presence of two merging CME-driven shocks. As they converge, a large-scale first-order Fermi acceleration process is facilitated by a ''macroscopic collapsing magnetic trap" (CMCT) process, similar to what happens in the case of shock waves in stream interacting regions \citet{2023Wijsen_b}. To our knowledge, no other study has addressed modeling of the energetic particle transport and acceleration in such a complex situation. For that, we use the modeling chain EUHFORIA-PARADISE, presented in Section~\ref{sec:models}. In Section~\ref{sec:EUHFORIA_res} we show the three different EUHFORIA background solar winds that were generated. The first two represent two isolated CMEs capable of accelerating energetic particles, and the third one corresponds to a combination of these CMEs, injected into the simulated domain such that they interact before arriving at 1~au. Such an interaction between a slow and a faster CME with overlapping trajectories was coined as ``cannibal CMEs" by \citet{2001Gopalswamy}. PARADISE results are presented in Section~\ref{sec:paradise_results}, where some of the more efficient particle acceleration mechanisms that appear in the presence of both CMEs are discussed, with a final summary presented in Section~\ref{sec:conclusions}.

\section{Models} \label{sec:models}

    The two models used for this study are the magnetohydrodynamic (MHD) model of solar wind and CMEs, EUHFORIA \citep{Pomoell2018}, and the particle acceleration and transport model, PARADISE \citep{Wijsen2019(1),2020WijsenPhDThesis}. EUHFORIA solves the ideal MHD  equations in three dimensions, to get a description of the solar wind and CME propagation in the inner heliosphere \citep[e.g.,][]{Verbeke2019,Scolini2019,Maharana2022}. 
    Next, PARADISE is used to evolve energetic particle distributions through the EUHFORIA-generated solar wind configurations, from 0.1~au. This is done by solving the focused transport equation \citep[FTE; e.g.,][]{1971Skilling,Isenberg1997,leRoux2009,2012leRouxWebb,2021vandenberg}. A description of the model and its setup used in this work are provided in Appendix~\ref{subsec:PARADISE}. As the simulation domain starts at 0.1~au, the results presented do not include the particle acceleration occurring in the corona.
    
    The modeling chain EUHFORIA-PARADISE has been used previously to successfully study and reproduce in-situ observed gradual SEP events \citep[e.g.,][]{2022Wijsen,2023Wijsen,2023Niemela}. 
    In this work, we take a different approach and study a synthetic solar eruptive event that consists of two fast, interacting, CMEs. These CMEs are modeled with the cone model \citep[e.g.,][]{2003xie,2013Millward}, and a uniform background solar wind is assumed (see Appendix~\ref{subsec:EUHFORIA}). To gain a comprehensive understanding of the impact of the interacting CMEs on the modeled SEP acceleration and transport, three distinct simulation set-ups are considered. In the first scenario, a single fast CME ($v_{inj} =$~1000~km~s$^{-1}$) is inserted in the heliospheric domain. In the second scenario, a very fast CME ($v_{inj} =$~2000$~\mathrm{km~s^{-1}}$) is inserted. Finally, in the third scenario, both CMEs are inserted 15.5 hours apart, to ensure that the shock of the fast CME merges with the shock of the slower CME just before they arrive at 1~au. In all cases, the CMEs are inserted from the inner boundary ($r = 0.1$~au) at the same position (0$^{\circ}$ in latitude and longitude) and with the same density ($10^{-17}~\mathrm{kg~m^{-3}}$), temperature (0.8~MK), and half-width (50$^{\circ}$).

    After the EUHFORIA simulations are obtained, a shock tracing model is used to extract the shocks' positions and properties. The algorithm details and performance are discussed in Appendix~\ref{sec:shock_tracing}. The output of the shock tracing model is used to continuously inject a mono-energetic population of 50~keV protons at the CME-driven shock. The injection intensity is scaled proportional to the local upstream solar wind density, resulting in a lower (higher) injection of protons when the CME shock propagates through regions of low (high) solar wind density. The parallel mean free path ($\lambda_{\parallel}$) is assumed to be 0.3~au for a 1~MeV proton, and we assume the perpendicular mean free path ($\lambda_{\perp}$) to scale proportional to the particle's gyroradius and $\lambda_{\parallel}$ \citep{2010Dröge,Wijsen2019(1)}. More details about PARADISE can be found Appendix~\ref{subsec:PARADISE}.

\section{RESULTS} \label{sec:Results}
\subsection{EUHFORIA results}\label{sec:EUHFORIA_res}
    
\begin{figure*}
    \centering
        \begin{interactive}{animation}{Movie.mp4}
        \includegraphics[width=0.65\textwidth]{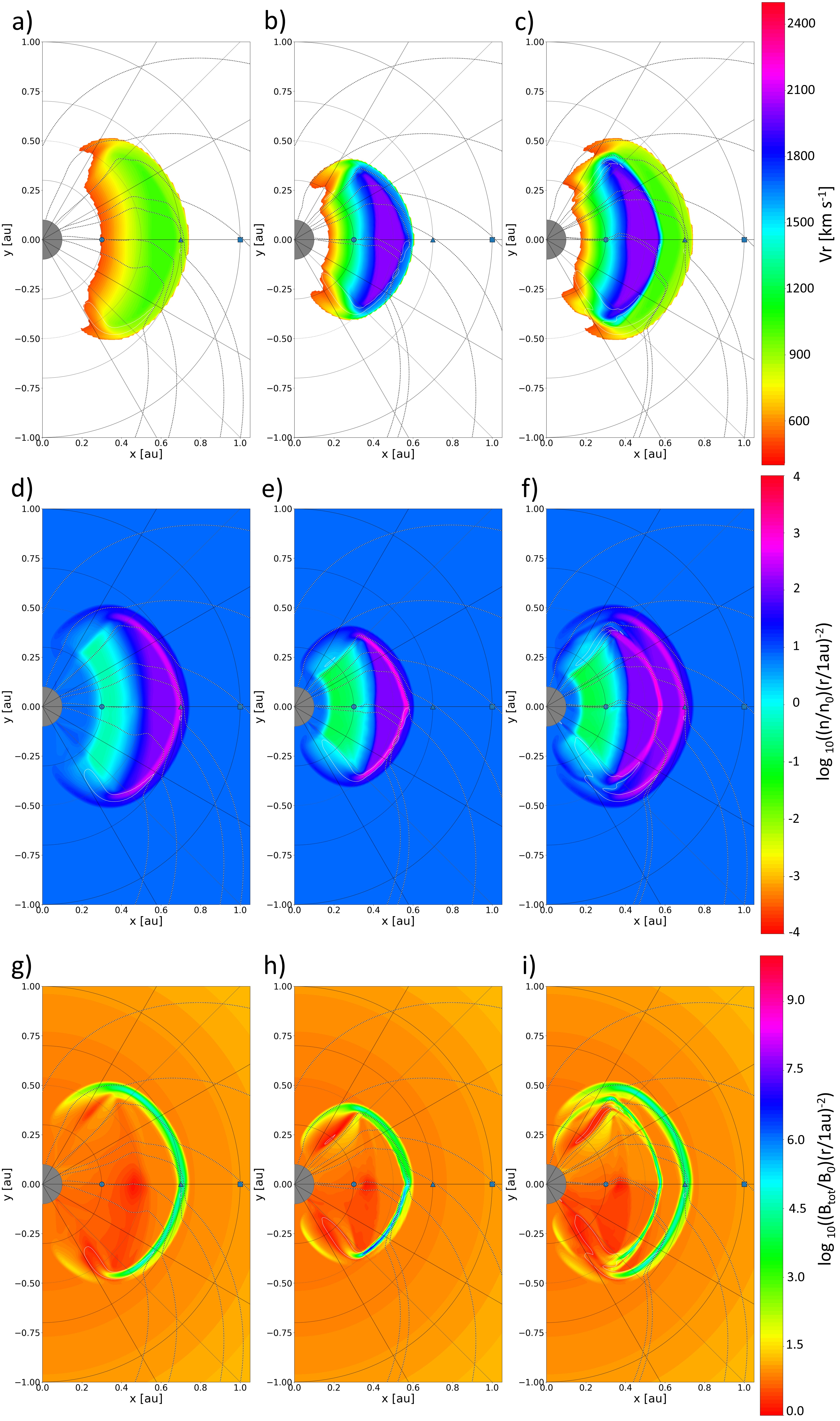}
    \end{interactive}
     \caption{The temporal evolution of the EUHFORIA simulation results for the three different cases (movie available as online additional material). The case with the isolated slower CME is shown on the left column panels, the case with the isolated faster CME is shown on the middle column panels and the case with both CMEs is shown on the right column panels. Panels~a), b), and c) show the radial solar wind speed above 500~km~s$^{-1}$}. Panels~d), e), and f) show the base 10 logarithm of EUHFORIA density (scaled with the square of the radial distance). Panels~g), h), and i) show the base 10 logarithm of EUHFORIA total magnetic field (scaled with the square of the radial distance). Black and white dashed lines show the projection to the equatorial plane of the magnetic field lines that connect the different observers to the inner boundary of EUHFORIA. Extra field lines are presented to showcase behavior at the flanks of the CMEs.The total duration of movie is 19 seconds.
    \label{fig:Solar_wind_sim}
    \end{figure*}

    Figure~\ref{fig:Solar_wind_sim} shows a snapshot of the three EUHFORIA simulations of the background solar wind. The solar wind speed, density, and magnetic field intensity are presented in the top, middle, and bottom panels, respectively. The latter two quantities are expressed in base 10 logarithm and scaled with the square of the radial distance. Panels~a), d), and g) correspond to the simulation with only the 1000~km~s$^{-1}$ CME (slower CME hereafter) inserted into the domain, whereas panels~b), e) and h) correspond to the simulation with only the 2000~km~s$^{-1}$ CME (faster CME hereafter). Panels~c), f), and i) correspond to the simulation with both CMEs inserted into the domain.

    The figure illustrates that, when both CMEs are present in the simulation, the faster CME is only slightly deformed when compared to the run with no slower CME. Moreover, due to the lower density in front of the faster CME, its shock wave propagates faster than the CME core. This effect can be seen more clearly in panels f) and i) of movie provided in the online material, and at the CME flanks in panels~f) and i) of Figure~\ref{fig:Solar_wind_sim}. At this point of the simulation, the flanks of the shock of the faster CME are about to start interacting with the shock sheath of the slower CME, while the CME nose is further away from it. We study the SEP events observed by virtual spacecraft placed at 0.3, 0.7, and 1~au from the Sun and along the main propagation direction of the CMEs (in Fig.~\ref{fig:Solar_wind_sim} blue circle, triangle, and square, respectively). We also plot a few interplanetary magnetic field (IMF) lines (white-black dashed lines), with a focus on those that connect the different spacecraft with the inner boundary of EUHFORIA. When only one CME is present in the domain, the connection to the shock follows a nominal (Parker) magnetic connection. On the contrary, when both CMEs are present, the slower CME drastically modifies the magnetic connection of the observers to the shock of the faster CME.

    Given the way the CMEs are inserted into the simulation domain, the observers establish a connection to the CME at different times. As the CME expands and propagates through the heliospheric domain, the observers connect with the west flank of the CME, and gradually the connection (the cobpoint) moves towards the CME nose, right as the CME-driven shock is at the position of the observer. Initially, the observer is magnetically connected to a quasi-perpendicular region of the shock. As the cobpoint moves towards the nose of the shock the shock obliquity becomes quasi-parallel.

    In the background solar wind scenario with both CMEs within the domain, the magnetic connection to the slower CME shock remains nominal. However, the presence of the slower CME between the observer and the faster CME results in a significant alteration of the magnetic connection to the faster CME. At the time of injection of the faster CME, the slower CME is close to 0.5~au. The predominant radial magnetic field topology in the wake of the slower CME causes the observer at 0.3~au to establish a good initial connection to the west flank of the faster CME-driven shock as soon as the faster CME is inserted into the domain. This connection moves rapidly towards the nose. As the faster CME reaches 0.3~au, it starts interacting with the trailing edge of the slower CME. By this time, the 0.7~au and 1~au observer establish a magnetic connection to the very western edge of the faster CME-driven shock, through the slower CME. Similarly to what happens when there is only one CME, the magnetic connection to the slower CME shock moves gradually towards its nose. However, the connection of the observers to the faster CME-driven shock remains fixed to a position at the western flank of the shock, until the observer is embedded downstream of the slower CME shock (see complementary material for Fig.~\ref{fig:Solar_wind_sim}). This is evident for the 0.7~au observer, where as soon as the slower CME reaches this position, the magnetic connection to the faster CME-driven shock moves rapidly toward its nose. On the other hand, the connection of the 1~au observer to the faster CME shock does not change until both shocks merge and arrive at its position.

\subsection{PARADISE results}
\label{sec:paradise_results}
    
\begin{figure*}[ht]
    \centering
    \includegraphics[width=0.98\textwidth]{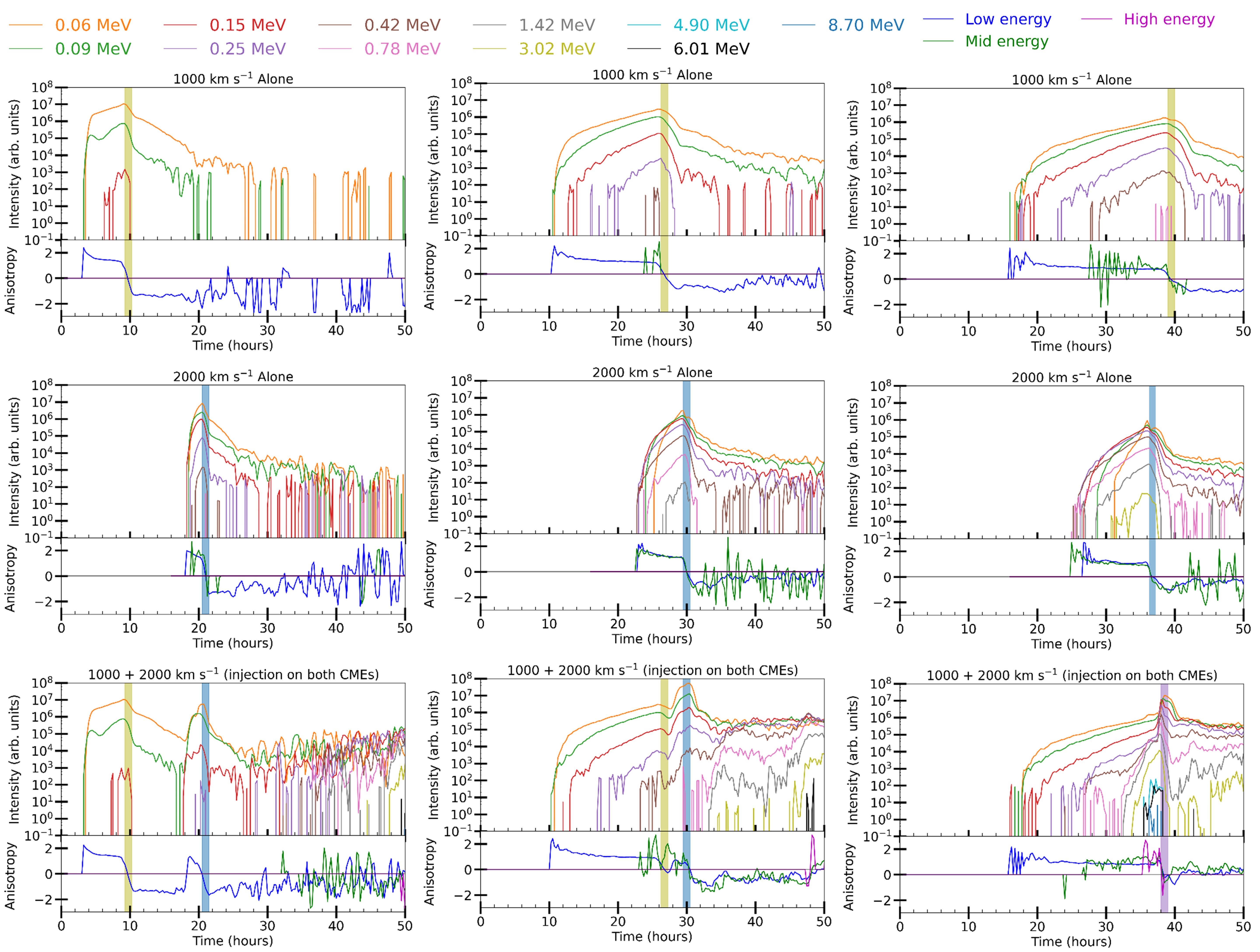}
    \caption{PARADISE simulations results for the three EUHFORIA simulations (different rows) at 0.3~au (panels~a, d, and, g), 0.7~au (panels~b, e, and h), and 1~au (panels~c, f, and i). Each panel consists of omnidirectional intensity-time profiles (top) and first-order parallel anisotropy time profiles (bottom).  Low, mid, and high energy anisotropy-time profiles are calculated using the lowest 4, middle 4, and highest 4 energy channels, respectively. Vertical shaded areas correspond to a time window of $\pm$30~minutes around the modeled shock.
    \label{fig:time_series}}
    \end{figure*}

    Figure~\ref{fig:time_series} shows the omnidirectional intensity and first-order parallel anisotropy time profiles for different proton energy channels registered by the three virtual spacecraft: 0.3~au (panels~a, d, and, g), 0.7~au (panels~b, e, and h), and 1~au (panels~c, f, and i). The SEP events are shown for the slower CME case (top row), the faster CME case (middle), and both CMEs included in the EUHFORIA simulation (bottom). Simulations where both CMEs are inserted and particles are injected at either the slower or the faster CME shock are presented in Appendix~\ref{sec:complementary_PARADISE} for completeness. The colored vertical bars in Fig.~\ref{fig:time_series} correspond to the $\pm$30~minutes time window centered around the time of the CME-driven shock passage over the observer.
    
    Profiles for the cases when only one CME is included are smooth and qualitatively similar. That is, in both cases the peak intensities and a sign switch in anisotropy are registered when the shock crosses the observer. The further into the heliosphere the observer is, the higher energy SEPs it registers, which is the result of continuous particle acceleration at the CME-driven shock\footnote{This is partly because our simulations start at 0.1~au and hence do not include the efficient particle acceleration processes that happen in the corona.}.  The faster CME is more efficient at accelerating particles than the slower CME at all radial distances, which is an expected consequence of its larger speed and stronger shock \citep[e.g.,~][]{1983Drury}.

\subsubsection{SEP events Characteristics at 0.3~au }
\label{sect:Charact_at_0_3au}

  The scenario which includes both CMEs in the EUHFORIA simulation (bottom row of Fig.~\ref{fig:time_series}), exhibits distinct characteristics compared to the single-CME configurations. At 0.3~au (Fig.~\ref{fig:time_series} panel~g), the intensity profiles reveal two separate peaks, each associated with a CME-driven shock. As expected, the onset time of the SEP event and the first intensity peak closely resemble those of the scenario featuring only the slower CME. However, the second intensity peak, attributed to the faster CME, differs notably from the peaks observed when only the faster CME is considered. Specifically, the second peak is somewhat lower than the one in Fig.~\ref{fig:time_series} panel~d), and its energy spectrum is considerably softer. In particular, when only the faster CME is present, particles are accelerated up to 0.42~MeV at 0.3~au, whereas in the presence of both CMEs (panel~g), the maximum energy reached when the faster CME reaches 0.3~au is only 0.15~MeV. This difference arises from two main factors: Firstly, as the faster CME travels in the wake of the slower CME, it encounters a less dense solar wind plasma, resulting consequently in fewer particles injected at the shock. Secondly, the speed differential across the shock of the faster CME is diminished when it propagates through the wake of the slower CME, leading to a decreased acceleration efficiency. Furthermore, other parameters affecting the particle acceleration rate (e.g., the shock obliquity or the ratio of the magnetic field intensity) undergo modifications, due to the different solar winds simulated, which could thus also contribute to the observed differences.

\subsubsection{SEP events Characteristics at 0.7~au }
\label{sect:Charact_at_0_7au}

     As described in Section~\ref{sec:EUHFORIA_res}, by the time the slower CME-driven shock is at 0.7~au (along the CMEs' central injection location line), the flanks of both shocks are close to each other, while at the nose, the shock driven by faster CME still has to enter the denser core of the slower CME, before arriving at its shock. Therefore, 0.7~au can be considered as the distance just before the shocks merge. For the observer at 0.7~au (panels~b, e, and h of Fig~\ref{fig:time_series}) the maximum energetic particle intensities for all cases are registered within a 5-hour window from each other, instead of the $\sim 12$~hours at 0.3~au. The profiles shown in panels~b) and h) exhibit a similar pattern until the passage of the slower CME's shock, when there is a switch in sign for the low-energy anisotropy. At this time, 26~hours after the injection of the slower CME, the peak intensities are very similar and the maximum energy of the registered protons falls in the 0.42~MeV energy channel. This is an expected result, as the upstream of the slower CME is always the unperturbed background solar wind. 
     Once the shock of the slower CME has passed the observer, the intensities decrease for about two hours across all energy levels. In the case involving both CMEs, intensities begin to increase again approximately 2.5 hours before the arrival of the faster CME's shock at 0.7~au, coinciding with the low-energy anisotropy switching back to positive values.  This indicates that the registered particles are mostly streaming upstream from this shock. Additionally, since most of the middle-energy protons are emitted by the faster-CME shock, their anisotropy only switches to negative values upon the arrival of this shock at the observer.
     
     All intensities keep increasing until the passage of the faster CME shock. Around this time, the intensities for the two lower energy channels (in panel~h) are one order of magnitude higher than in the case with the faster CME alone (panel~e). 
     This increase can be attributed to the CME-driven shock starting to propagate inside the denser core of the preceding, slower, CME from ${\sim}$0.5 au onward. This results in a larger injection of 50~keV particles at the shock front. However, it is important to note that the shock's strength, as indicated by its Alfvén Mach number (not shown), decreases when the faster CME shock traverses the slower CME compared to when no preceding CME is present. This decrease in shock strength occurs because the slower CME provides a denser medium through which the faster CME has to propagate, leading to a more pronounced deceleration of the faster CME. This reduction in shock speed is substantial enough to offset the decrease in Alfv\'en speed (due to the enhanced density), resulting in the shock having a lower Alfv\'en Mach number. The weaker shock explains why the maximum energy achieved by protons in the scenario involving both CMEs (0.78 MeV) is lower than that in the case of the faster CME alone (1.42 MeV).

    \subsubsection{SEP events Characteristics at 1~au }
    \label{sect:Charact_at_1au}
    At 1~au, since both shocks have already merged, there is no longer a distinction between the shocks driven by the slower and faster CMEs. Therefore, only one switch in the sign of anisotropies can be observed. Similarly to the other positions, the onset time of the SEP event at 1~au in panel~i) coincides with panel~c), as particles at the beginning travel in the unperturbed medium. However, the onset of the 0.42~MeV channel is approximately 3~hours earlier in the bottom panel, which means that these particles are mainly being generated at the faster CME-driven shock, propagate through the slower CME, and arrive at 1~au. However, higher energy particles do not show the same behavior, as they are not generated until later in the simulation. 
    As the faster shock approaches the sheath of the slower shock, the magnetic amplification of the slower CME's sheath acts as a magnetic mirror, redirecting particles towards the faster CME shock. Similarly, the magnetic enhancement of the faster shock also serves as a magnetic mirror. The convergence of these shock waves thus resembles a collapsing magnetic trap, facilitating an efficient large-scale first-order Fermi acceleration process. This phenomenon parallels the case examined in \citet{2023Wijsen_b} for interacting SIR shock waves, termed the ``macroscopic collapsing magnetic trap'' (MCMT). It is worth noting that in addition to the acceleration associated with this MCMT, particles also undergo acceleration by scattering across either shock wave, representing the standard first-order Fermi acceleration process occurring at shocks \citep[e.g.,][]{1983Drury}. 
    As the faster shock approaches the slower shock, the MCMT process contributes to the acceleration of particles up to higher energies. These particles next serve as a seed population for the stronger fully merged shock, which is capable of accelerating them to even higher energies. This efficient acceleration process results in the highest energy particles to be registered in the 8.70~MeV channel when the shock reaches the observer at 1 au (panel~f). For comparison, the highest energy particles registered for the case with only the slower CME (panel~c) are within the 0.42~MeV energy channel, and for the simulation with only the faster CME (panel~f) the highest energy particles fall in the 3.02~MeV energy channel.

\section{Summary and Conclusions} \label{sec:conclusions}

 In this study, we investigated the SEP acceleration in the interplanetary space induced by single and interacting CMEs. We employed three different simulation set-ups for EUHFORIA runs. The first two scenarios considered the injection of a single cone CME with a velocity of 1000~km~s$^{-1}$ (slower CME), and a single cone CME with a velocity of 2000~km~s$^{-1}$ (faster CME). The third scenario included the injection of both CMEs 15.5~hours apart. In this way, the slower CME is overtaken by the faster CME between 0.7~au and 1~au (in the direction along the CMEs' central injection location line). This interaction causes the two CME-driven shocks to be indistinguishable at 1~au. After the EUHFORIA simulations are completed, an automatic algorithm extracts the position and characteristics of each shock, which is subsequently used by PARADISE to inject particles continuously at the shock front. To understand the particle acceleration in the interplanetary space under such complex conditions, a 50~keV mono-energetic proton population, proportional to the upstream wind density, was considered. We analyze the resulting energetic particle events registered at three virtual spacecraft located at 0.3~au, 0.7~au, and 1~au along the direction the CMEs are launched. 
 
 In simulations featuring only one CME, particles eventually propagate upstream into the uniform background solar wind after undergoing acceleration at the CME-driven shock. However, in the scenario with two CMEs, particles injected at the shock of the faster CME escape into the perturbed wake of the slower and eventually reach the sheath of the slower CME. The results of our study show that:

    \begin{itemize}
        \item Magnetic connectivity to the faster CME changes substantially when the slower CME is included in the simulation. 
        \item At smaller radial distances and before CME-driven shocks start interacting, particle acceleration by the faster CME is less efficient when both CMEs are present in the simulation, as CME-driven shock properties are strongly modified in the wake of the slower CME.
        \item A large fraction of the particles injected at the faster CME's shock is trapped between the downstream of the slower CME-driven shock and the upstream of the faster CME-driven shock. These particles suffer the effects of a MCMT, which accelerates the particles alongside the first-order Fermi acceleration process occurring at each shock wave.
        Hence, particles are subjected to multiple acceleration processes, resulting in a more energetic SEP event. The MCMT effect is effective when the faster CME shock is close to the enhanced magnetic field downstream of the slower CME shock.
        \item The merged shock is stronger than each shock, and further accelerates the particles to higher energies. 
      
    \end{itemize}
    
   Observational studies have already highlighted the importance of the preconditioning of the solar wind during some of the most intense SEP events \citep{2004Gopalswamy,2020Zhuang}. This study complements the results of \citet{2012Li}, where authors discuss how a preceding CME may enhance turbulence and hence give rise to more efficient particle acceleration by a subsequent CME. In contrast, our investigation focuses on understanding the large-scale processes influencing energetic particle acceleration and transport amidst interacting CMEs.
   
   The analysis presented in this work is facilitated by employing two advanced models, EUHFORIA and PARADISE. However, ongoing improvements are necessary to better capture the complexities involving interacting CMEs. For instance, the CME models used in this work do not include the inner magnetic field structure of CMEs. Including a magnetized CME may modify the IMF and the particle behavior \citep[e.g.,][]{2023Niemela}. This aspect will be explored in a follow-up study using the linear force-free (LFF) Spheromak CME model \citep{Verbeke2019}, Flux Rope in 3D \citep[FRi3D,][]{2016Isavnin,Maharana2022} CME model, which are already implemented into EUHFORIA. Additionally, efforts are underway to couple EUHFORIA with COCONUT, a 3D MHD model for the solar corona \citep[see][]{2022Perri,2023Perri}, allowing us to extend our investigation to the corona, where particle acceleration is expected to be particularly efficient \citep[][and references therin]{1993Kallenrode,2003Rice,2016Desai}. These results can also be complemented by using data from Parker Solar Probe or Solar Orbiter.

\section*{Acknowledgments}
EUHFORIA was created as a joint effort between KU Leuven and the University of Helsinki and was developed further by the project EUHFORIA 2.0, a European Union's Horizon 2020 research and innovation program under grant agreement No 870405 and the ESA project 'Heliospheric modelling techniques' (Contract No. 4000133080/20/NL/CRS). These results were also obtained in the framework of the projects C14/19/089 (C1 project Internal Funds KU Leuven), G.0B58.23N and G.0025.23N (FWO-Vlaanderen), SIDC Data Exploitation (ESA Prodex-12), and BelSPo project B2/191/P1/SWiM. The ROB team thanks the Belgian Federal Science Policy Office (BELSPO) for the provision of financial support in the framework of the PRODEX Programme of the European Space Agency (ESA) under contract numbers 4000134088, 4000112292, 4000134474, and 4000136424. J.M. also acknowledges the funding by the FEDtWin project PERIHELION.
N.W.\ acknowledges support from the Research Foundation - Flanders (FWO-Vlaanderen, fellowship no.\ 1184319N).
A.A. acknowledges the project PID2022-136828NB-C41 that received financial support from
MICIU/AEI/10.13039/501100011033 and FEDER, UE; the support through the
``Center of Excellence Mar\'{i}a de Maeztu 2020-2023'' award to the ICCUB (CEX2019-000918-M); and, from the Departament de Recerca i
Universitats of Generalitat de Catalunya through grant 2021SGR00679.
Computational resources and services used in this work were provided by the VSC (Flemish Supercomputer Centre), funded by the FWO and the Flemish Government-Department EWI.

\bibliography{Bibliography}{}

\begin{thebibliography}{}
\expandafter\ifx\csname natexlab\endcsname\relax\def\natexlab#1{#1}\fi
\providecommand{\url}[1]{\href{#1}{#1}}
\providecommand{\dodoi}[1]{doi:~\href{http://doi.org/#1}{\nolinkurl{#1}}}
\providecommand{\doeprint}[1]{\href{http://ascl.net/#1}{\nolinkurl{http://ascl.net/#1}}}
\providecommand{\doarXiv}[1]{\href{https://arxiv.org/abs/#1}{\nolinkurl{https://arxiv.org/abs/#1}}}

\bibitem[{{Agueda} \& {Vainio}(2013)}]{2013Agueda}
{Agueda}, N., \& {Vainio}, R. 2013, Journal of Space Weather and Space Climate, 3, A10, \dodoi{10.1051/swsc/2013034}

\bibitem[{{Axford} {et~al.}(1977){Axford}, {Leer}, \& {Skadron}}]{1977Axford}
{Axford}, W.~I., {Leer}, E., \& {Skadron}, G. 1977, in International Cosmic Ray Conference, Vol.~11, International Cosmic Ray Conference, 132

\bibitem[{Bell(1978)}]{1978Bell}
Bell, A.~R. 1978, Monthly Notices of the Royal Astronomical Society, 182, 147, \dodoi{10.1093/mnras/182.2.147}

\bibitem[{{Blandford} \& {Ostriker}(1978)}]{1978Blandford}
{Blandford}, R.~D., \& {Ostriker}, J.~P. 1978, \apjl, 221, L29, \dodoi{10.1086/182658}

\bibitem[{Borovikov {et~al.}(2018)Borovikov, Sokolov, Roussev, Taktakishvili, \& Gombosi}]{2018Borovikov}
Borovikov, D., Sokolov, I.~V., Roussev, I.~I., Taktakishvili, A., \& Gombosi, T.~I. 2018, The Astrophysical Journal, 864, 88, \dodoi{10.3847/1538-4357/aad68d}

\bibitem[{{Desai} \& {Giacalone}(2016)}]{2016Desai}
{Desai}, M., \& {Giacalone}, J. 2016, Living Reviews in Solar Physics, 13, 3, \dodoi{10.1007/s41116-016-0002-5}

\bibitem[{{Ding} {et~al.}(2022){Ding}, {Wijsen}, {Li}, \& {Poedts}}]{2022Ding_b}
{Ding}, Z., {Wijsen}, N., {Li}, G., \& {Poedts}, S. 2022, \aap, 668, A71, \dodoi{10.1051/0004-6361/202244732}

\bibitem[{{Drury}(1983)}]{1983Drury}
{Drury}, L.~O. 1983, Reports on Progress in Physics, 46, 973, \dodoi{10.1088/0034-4885/46/8/002}

\bibitem[{Dröge {et~al.}(2010)Dröge, Kartavykh, Klecker, \& Kovaltsov}]{2010Dröge}
Dröge, W., Kartavykh, Y.~Y., Klecker, B., \& Kovaltsov, G.~A. 2010, The Astrophysical Journal, 709, 912, \dodoi{10.1088/0004-637X/709/2/912}

\bibitem[{{Gopalswamy} {et~al.}(2001){Gopalswamy}, {Yashiro}, {Kaiser}, {Howard}, \& {Bougeret}}]{2001Gopalswamy}
{Gopalswamy}, N., {Yashiro}, S., {Kaiser}, M.~L., {Howard}, R.~A., \& {Bougeret}, J.~L. 2001, \apjl, 548, L91, \dodoi{10.1086/318939}

\bibitem[{{Gopalswamy} {et~al.}(2004){Gopalswamy}, {Yashiro}, {Krucker}, {Stenborg}, \& {Howard}}]{2004Gopalswamy}
{Gopalswamy}, N., {Yashiro}, S., {Krucker}, S., {Stenborg}, G., \& {Howard}, R.~A. 2004, Journal of Geophysical Research (Space Physics), 109, A12105, \dodoi{10.1029/2004JA010602}

\bibitem[{{Hasselmann} \& {Wibberenz}(1970)}]{1970Hasselmann}
{Hasselmann}, K., \& {Wibberenz}, G. 1970, \apj, 162, 1049, \dodoi{10.1086/150736}

\bibitem[{{Isavnin}(2016)}]{2016Isavnin}
{Isavnin}, A. 2016, \apj, 833, 267, \dodoi{10.3847/1538-4357/833/2/267}

\bibitem[{Isenberg(1997)}]{Isenberg1997}
Isenberg, P.~A. 1997, Journal of Geophysical Research: Space Physics, 102, 4719, \dodoi{10.1029/96ja03671}

\bibitem[{{Jokipii}(1966)}]{1966Jokipii}
{Jokipii}, J.~R. 1966, \apj, 146, 480, \dodoi{10.1086/148912}

\bibitem[{{Kallenrode} {et~al.}(1993){Kallenrode}, {Wibberenz}, {Kunow}, {M{\"u}ller-Mellin}, {Stolpovskii}, \& {Kontor}}]{1993Kallenrode}
{Kallenrode}, M.~B., {Wibberenz}, G., {Kunow}, H., {et~al.} 1993, \solphys, 147, 377, \dodoi{10.1007/BF00690726}

\bibitem[{le~Roux \& Webb(2009)}]{leRoux2009}
le~Roux, J.~A., \& Webb, G.~M. 2009, The Astrophysical Journal, 693, 534, \dodoi{10.1088/0004-637x/693/1/534}

\bibitem[{{le Roux} \& {Webb}(2012)}]{2012leRouxWebb}
{le Roux}, J.~A., \& {Webb}, G.~M. 2012, \apj, 746, 104, \dodoi{10.1088/0004-637X/746/1/104}

\bibitem[{{Li} {et~al.}(2012){Li}, {Moore}, {Mewaldt}, {Zhao}, \& {Labrador}}]{2012Li}
{Li}, G., {Moore}, R., {Mewaldt}, R.~A., {Zhao}, L., \& {Labrador}, A.~W. 2012, \ssr, 171, 141, \dodoi{10.1007/s11214-011-9823-7}

\bibitem[{{Lugaz} {et~al.}(2017){Lugaz}, {Temmer}, {Wang}, \& {Farrugia}}]{2017Lugaz}
{Lugaz}, N., {Temmer}, M., {Wang}, Y., \& {Farrugia}, C.~J. 2017, \solphys, 292, 64, \dodoi{10.1007/s11207-017-1091-6}

\bibitem[{{Luhmann} {et~al.}(2007){Luhmann}, {Ledvina}, {Krauss-Varban}, {Odstrcil}, \& {Riley}}]{2007Luhmann}
{Luhmann}, J.~G., {Ledvina}, S.~A., {Krauss-Varban}, D., {Odstrcil}, D., \& {Riley}, P. 2007, Advances in Space Research, 40, 295, \dodoi{10.1016/j.asr.2007.03.089}

\bibitem[{{Luhmann} {et~al.}(2010){Luhmann}, {Ledvina}, {Odstrcil}, {Owens}, {Zhao}, {Liu}, \& {Riley}}]{2010Luhmann}
{Luhmann}, J.~G., {Ledvina}, S.~A., {Odstrcil}, D., {et~al.} 2010, Advances in Space Research, 46, 1, \dodoi{10.1016/j.asr.2010.03.011}

\bibitem[{{Luhmann} {et~al.}(2017){Luhmann}, {Mays}, {Odstrcil}, {Li}, {Bain}, {Lee}, {Galvin}, {Mewaldt}, {Cohen}, {Leske}, {Larson}, \& {Futaana}}]{2017Luhmann}
{Luhmann}, J.~G., {Mays}, M.~L., {Odstrcil}, D., {et~al.} 2017, Space Weather, 15, 934, \dodoi{10.1002/2017SW001617}

\bibitem[{{Maharana} {et~al.}(2022){Maharana}, {Isavnin}, {Scolini}, {Wijsen}, {Rodriguez}, {Mierla}, {Magdaleni{\'c}}, \& {Poedts}}]{Maharana2022}
{Maharana}, A., {Isavnin}, A., {Scolini}, C., {et~al.} 2022, Advances in Space Research, 70, 1641, \dodoi{10.1016/j.asr.2022.05.056}

\bibitem[{{Millward} {et~al.}(2013){Millward}, {Biesecker}, {Pizzo}, \& {de Koning}}]{2013Millward}
{Millward}, G., {Biesecker}, D., {Pizzo}, V., \& {de Koning}, C.~A. 2013, Space Weather, 11, 57, \dodoi{10.1002/swe.20024}

\bibitem[{{Niemela} {et~al.}(2023){Niemela}, {Wijsen}, {Aran}, {Rodriguez}, {Magdalenic}, \& {Poedts}}]{2023Niemela}
{Niemela}, A., {Wijsen}, N., {Aran}, A., {et~al.} 2023, \aap, 679, A93, \dodoi{10.1051/0004-6361/202347116}

\bibitem[{Odstrcil(2003)}]{2003ODSTRCIL}
Odstrcil, D. 2003, Advances in Space Research, 32, 497, \dodoi{https://doi.org/10.1016/S0273-1177(03)00332-6}

\bibitem[{{Pal} {et~al.}(2023){Pal}, {Balmaceda}, {Weiss}, {Nieves-Chinchilla}, {Carcaboso}, {Kilpua}, \& {M{\"o}stl}}]{2023Pal}
{Pal}, S., {Balmaceda}, L., {Weiss}, A.~J., {et~al.} 2023, Frontiers in Astronomy and Space Sciences, 10, 1195805, \dodoi{10.3389/fspas.2023.1195805}

\bibitem[{{Palmerio} {et~al.}(2022){Palmerio}, {Lee}, {Mays}, {Luhmann}, {Lario}, {S{\'a}nchez-Cano}, {Richardson}, {Vainio}, {Stevens}, {Cohen}, {Steinvall}, {M{\"o}stl}, {Weiss}, {Nieves-Chinchilla}, {Li}, {Larson}, {Heyner}, {Bale}, {Galvin}, {Holmstr{\"o}m}, {Khotyaintsev}, {Maksimovic}, \& {Mitrofanov}}]{2022Palmerio}
{Palmerio}, E., {Lee}, C.~O., {Mays}, M.~L., {et~al.} 2022, Space Weather, 20, e2021SW002993, \dodoi{10.1029/2021SW002993}

\bibitem[{{Perri} {et~al.}(2023){Perri}, {Ku{\'z}ma}, {Brchnelova}, {Baratashvili}, {Zhang}, {Leitner}, {Lani}, \& {Poedts}}]{2023Perri}
{Perri}, B., {Ku{\'z}ma}, B., {Brchnelova}, M., {et~al.} 2023, \apj, 943, 124, \dodoi{10.3847/1538-4357/ac9799}

\bibitem[{{Perri} {et~al.}(2022){Perri}, {Leitner}, {Brchnelova}, {Baratashvili}, {Ku{\'z}ma}, {Zhang}, {Lani}, \& {Poedts}}]{2022Perri}
{Perri}, B., {Leitner}, P., {Brchnelova}, M., {et~al.} 2022, \apj, 936, 19, \dodoi{10.3847/1538-4357/ac7237}

\bibitem[{{Pomoell} \& {Poedts}(2018)}]{Pomoell2018}
{Pomoell}, J., \& {Poedts}, S. 2018, Journal of Space Weather and Space Climate, 8, A35, \dodoi{10.1051/swsc/2018020}

\bibitem[{{Reames}(1999)}]{1999Reames}
{Reames}, D.~V. 1999, \ssr, 90, 413, \dodoi{10.1023/A:1005105831781}

\bibitem[{Rice {et~al.}(2003)Rice, Zank, \& Li}]{2003Rice}
Rice, W. K.~M., Zank, G.~P., \& Li, G. 2003, Journal of Geophysical Research: Space Physics, 108, \dodoi{https://doi.org/10.1029/2002JA009756}

\bibitem[{{Schrijver} {et~al.}(2015){Schrijver}, {Kauristie}, {Aylward}, {Denardini}, {Gibson}, {Glover}, {Gopalswamy}, {Grande}, {Hapgood}, {Heynderickx}, {Jakowski}, {Kalegaev}, {Lapenta}, {Linker}, {Liu}, {Mandrini}, {Mann}, {Nagatsuma}, {Nandy}, {Obara}, {Paul O'Brien}, {Onsager}, {Opgenoorth}, {Terkildsen}, {Valladares}, \& {Vilmer}}]{2015Schrijver}
{Schrijver}, C.~J., {Kauristie}, K., {Aylward}, A.~D., {et~al.} 2015, Advances in Space Research, 55, 2745, \dodoi{10.1016/j.asr.2015.03.023}

\bibitem[{{Scolini} {et~al.}(2019){Scolini}, {Rodriguez}, {Mierla}, {Pomoell}, \& {Poedts}}]{Scolini2019}
{Scolini}, C., {Rodriguez}, L., {Mierla}, M., {Pomoell}, J., \& {Poedts}, S. 2019, \aap, 626, A122, \dodoi{10.1051/0004-6361/201935053}

\bibitem[{Scolini {et~al.}(2020)Scolini, Chan{\'{e}}, Temmer, Kilpua, Dissauer, Veronig, Palmerio, Pomoell, Dumbovi{\'{c}}, Guo, Rodriguez, \& Poedts}]{Scolini2020}
Scolini, C., Chan{\'{e}}, E., Temmer, M., {et~al.} 2020, The Astrophysical Journal Supplement Series, 247, 21, \dodoi{10.3847/1538-4365/ab6216}

\bibitem[{{Shen} {et~al.}(2017){Shen}, {Wang}, {Shen}, \& {Feng}}]{2017Feng}
{Shen}, F., {Wang}, Y., {Shen}, C., \& {Feng}, X. 2017, \solphys, 292, 104, \dodoi{10.1007/s11207-017-1129-9}

\bibitem[{{Skilling}(1971)}]{1971Skilling}
{Skilling}, J. 1971, \apj, 170, 265, \dodoi{10.1086/151210}

\bibitem[{{Telloni} {et~al.}(2021){Telloni}, {Scolini}, {M{\"o}stl}, {Zank}, {Zhao}, {Weiss}, {Reiss}, {Laker}, {Perrone}, {Khotyaintsev}, {Steinvall}, {Sorriso-Valvo}, {Horbury}, {Wimmer-Schweingruber}, {Bruno}, {D'Amicis}, {De Marco}, {Jagarlamudi}, {Carbone}, {Marino}, {Stangalini}, {Nakanotani}, {Adhikari}, {Liang}, {Woodham}, {Davies}, {Hietala}, {Perri}, {G{\'o}mez-Herrero}, {Rodr{\'\i}guez-Pacheco}, {Antonucci}, {Romoli}, {Fineschi}, {Maksimovic}, {Sou{\v{c}}ek}, {Chust}, {Kretzschmar}, {Vecchio}, {M{\"u}ller}, {Zouganelis}, {Winslow}, {Giordano}, {Mancuso}, {Susino}, {Ivanovski}, {Messerotti}, {O'Brien}, {Evans}, \& {Angelini}}]{2021Telloni}
{Telloni}, D., {Scolini}, C., {M{\"o}stl}, C., {et~al.} 2021, \aap, 656, A5, \dodoi{10.1051/0004-6361/202140648}

\bibitem[{{Temmer} {et~al.}(2014){Temmer}, {Veronig}, {Peinhart}, \& {Vr{\v{s}}nak}}]{2014Temmer}
{Temmer}, M., {Veronig}, A.~M., {Peinhart}, V., \& {Vr{\v{s}}nak}, B. 2014, \apj, 785, 85, \dodoi{10.1088/0004-637X/785/2/85}

\bibitem[{Thomas~Lewiner \& Tavares(2003)}]{2003Lewiner}
Thomas~Lewiner, Hélio~Lopes, A. W.~V., \& Tavares, G. 2003, Journal of Graphics Tools, 8, 1, \dodoi{10.1080/10867651.2003.10487582}

\bibitem[{{van den Berg} {et~al.}(2020){van den Berg}, {Strauss}, \& {Effenberger}}]{2020vandenberg}
{van den Berg}, J., {Strauss}, D.~T., \& {Effenberger}, F. 2020, \ssr, 216, 146, \dodoi{10.1007/s11214-020-00771-x}

\bibitem[{{van den Berg} {et~al.}(2021){van den Berg}, {Engelbrecht}, {Wijsen}, \& {Strauss}}]{2021vandenberg}
{van den Berg}, J.~P., {Engelbrecht}, N.~E., {Wijsen}, N., \& {Strauss}, R.~D. 2021, \apj, 922, 200, \dodoi{10.3847/1538-4357/ac2736}

\bibitem[{{Verbeke} {et~al.}(2019){Verbeke}, {Pomoell}, \& {Poedts}}]{Verbeke2019}
{Verbeke}, C., {Pomoell}, J., \& {Poedts}, S. 2019, \aap, 627, A111, \dodoi{10.1051/0004-6361/201834702}

\bibitem[{{Wang} {et~al.}(2019){Wang}, {Giacalone}, {Yan}, {Ding}, {Li}, {Lu}, \& {Shan}}]{2019Wang}
{Wang}, X., {Giacalone}, J., {Yan}, Y., {et~al.} 2019, \apj, 885, 66, \dodoi{10.3847/1538-4357/ab4655}

\bibitem[{{Wijsen}(2020)}]{2020WijsenPhDThesis}
{Wijsen}, N. 2020, PhD thesis, Katholieke University of Leuven (Belgium) and University of Barcelona (Spain)

\bibitem[{{Wijsen} {et~al.}(2019){Wijsen}, {Aran}, {Pomoell}, \& {Poedts}}]{Wijsen2019(1)}
{Wijsen}, N., {Aran}, A., {Pomoell}, J., \& {Poedts}, S. 2019, \aap, 622, A28, \dodoi{10.1051/0004-6361/201833958}

\bibitem[{{Wijsen} {et~al.}(2023{\natexlab{a}}){Wijsen}, {Li}, {Ding}, {Lario}, {Poedts}, {Filwett}, {Allen}, \& {Dayeh}}]{2023Wijsen_b}
{Wijsen}, N., {Li}, G., {Ding}, Z., {et~al.} 2023{\natexlab{a}}, Journal of Geophysical Research (Space Physics), 128, e2022JA031203, \dodoi{10.1029/2022JA031203}

\bibitem[{{Wijsen} {et~al.}(2022){Wijsen}, {Aran}, {Scolini}, {Lario}, {Afanasiev}, {Vainio}, {Sanahuja}, {Pomoell}, \& {Poedts}}]{2022Wijsen}
{Wijsen}, N., {Aran}, A., {Scolini}, C., {et~al.} 2022, \aap, 659, A187, \dodoi{10.1051/0004-6361/202142698}

\bibitem[{{Wijsen} {et~al.}(2023{\natexlab{b}}){Wijsen}, {Lario}, {S{\'a}nchez-Cano}, {Jebaraj}, {Dresing}, {Richardson}, {Aran}, {Kouloumvakos}, {Ding}, {Niemela}, {Palmerio}, {Carcaboso}, {Vainio}, {Afanasiev}, {Pinto}, {Pacheco}, {Poedts}, \& {Heyner}}]{2023Wijsen}
{Wijsen}, N., {Lario}, D., {S{\'a}nchez-Cano}, B., {et~al.} 2023{\natexlab{b}}, \apj, 950, 172, \dodoi{10.3847/1538-4357/acd1ed}

\bibitem[{Xie {et~al.}(2004)Xie, Ofman, \& Lawrence}]{2003xie}
Xie, H., Ofman, L., \& Lawrence, G. 2004, Journal of Geophysical Research: Space Physics, 109, \dodoi{https://doi.org/10.1029/2003JA010226}

\bibitem[{Young {et~al.}(2021)Young, Schwadron, Gorby, Linker, Caplan, Downs, Török, Riley, Lionello, Titov, Mewaldt, \& Cohen}]{2021Young}
Young, M.~A., Schwadron, N.~A., Gorby, M., {et~al.} 2021, The Astrophysical Journal, 909, 160, \dodoi{10.3847/1538-4357/abdf5f}

\bibitem[{Zank {et~al.}(2014)Zank, le~Roux, Webb, Dosch, \& Khabarova}]{Zank2014}
Zank, G.~P., le~Roux, J.~A., Webb, G.~M., Dosch, A., \& Khabarova, O. 2014, The Astrophysical Journal, 797, 28, \dodoi{10.1088/0004-637x/797/1/28}

\bibitem[{{Zhuang} {et~al.}(2020){Zhuang}, {Lugaz}, {Gou}, {Ding}, \& {Wang}}]{2020Zhuang}
{Zhuang}, B., {Lugaz}, N., {Gou}, T., {Ding}, L., \& {Wang}, Y. 2020, \apj, 901, 45, \dodoi{10.3847/1538-4357/abaef9}

\end{thebibliography}
\bibliographystyle{aasjournal}

\appendix
\section{EUFHORIA set up}\label{subsec:EUHFORIA}

    In this work, we use the heliospheric module of EUHFORIA which solves the ideal MHD equations from a heliocentric distance of $r=0.1$~au onward \citep[see][for detals]{Pomoell2018}. Moreover, we assume uniform inner boundary conditions. Specifically, we assume that the radial solar wind velocity is $v_r = 400$ km/s, the plasma number density is $n=7.3\times10^8~$m$^{-3}$, the thermal pressure is $P=3.3$~nPa, and the radial magnetic field strength is $B_{r} = 192$~nT. 
    Furthermore,  the resolution of EUHFORIA's simulation domain was configured to be 1$^{\circ}$ for both the longitudinal and latitudinal coordinates and 0.0039 AU for the radial coordinate. The outer boundary was set at 4~au.

\section{Shock tracing}\label{sec:shock_tracing}

     The shock tracing model described in Appendix~A of \citet{2022Wijsen} was used for locating the CME-driven shock position in the EUHFORIA 3D simulation domain. In summary, the shock tracing method uses the marching cube algorithm \citep{2003Lewiner} to locate the shock position in each of the EUHFORIA snapshots. Shock properties are calculated using the Rankine-Hugoniot relations. For the cases with only one CME injected into the domain, the procedure is straightforward, as no other transients are present in the simulation. In the first step, the code determines the divergence of the solar wind in the whole domain. The outermost position where the divergence is negative is considered the position of the shock.

\begin{figure*}[htp]
    \plotone{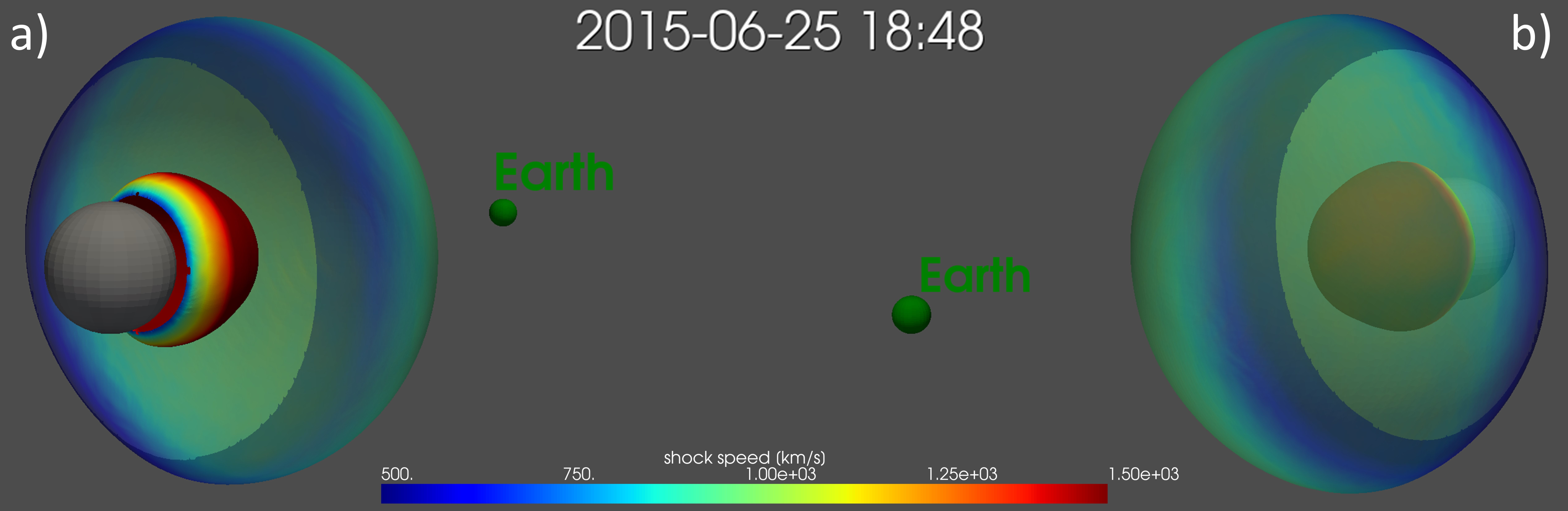}
    \caption{Visualization of the position of the shocks detected by the automatic shock tracing algorithm, corresponding to panel c) of Figure~\ref{fig:Solar_wind_sim}. The position of the Earth (not to scale) is marked with a green sphere, while the Sun is represented with a grey sphere. The gray sphere represents the inner boundary of EUHFORIA. a) Rear view. b) Front view.  
    \label{fig:shock_tracer}}
    \end{figure*}

    Similarly to the case described in \citet{2023Niemela}, having multiple CMEs that drive shocks complicates the detection of individual shocks. For a proper extraction of the faster CME shock location, when both CMEs are present in the domain, further post-processing of the EUHFORIA output was done. As the position of the slower CME is not affected by the faster CME injection and propagation (at least up to 0.7~au), the position of the slower shock was extracted similarly to the isolated case. Once the faster CME is injected, the snapshot from the isolated EUHFORIA simulation for the slower CME is subtracted from the snapshots corresponding to the EUHFORIA simulation with both CMEs. This forces the divergence generated by the slower CME to be approximately zero, and the process of obtaining the shock of the faster CME becomes analogous to the isolated case. Once the faster CME passes the position of the slower CME-driven shock, the algorithm only detects the position of the outermost shock, which corresponds to the merged shock. Figure~\ref{fig:shock_tracer} shows the position of the CME-driven shocks from two different viewing points. Using this information, the model also computes the properties of the shock which are eventually used to calculate the particle injection properties.

\section{PARADISE set up}\label{subsec:PARADISE}

    PARADISE evolves energetic particle distributions in EUHFORIA solar wind configurations by solving the focused transport equation \citep[FTE; e.g.,][]{1971Skilling,Isenberg1997,leRoux2009,2012leRouxWebb,Zank2014} stochastically using a time-forward Monte Carlo approach. The time-dependent FTE solved by PARADISE includes the effect of the solar wind turbulence through a set of phase-space diffusion processes \citep[for a recent review see e.g.,][]{2020vandenberg}. In \citet{Wijsen2019(1)} and Chapter~3 of \citet{2020WijsenPhDThesis} a detailed description of the PARADISE model is available.
    
     All the PARADISE runs are done by continuously injecting a 50~keV mono-energetic particle population at the CME-driven shock position, which is located using the model presented in Appendix~\ref{sec:shock_tracing}. The injection intensity is scaled proportional to the upstream solar wind density, as we assume that the seed population originates from the background solar wind \citep[see e.g.,][for similar approaches]{2022Ding_b}{}{}. This results in a lower (higher) injection of protons when the CME shock propagates through regions of low (high) solar wind density.

    The pitch-angle diffusion process included in the PARADISE simulations is derived from quasi-linear theory \citep{1966Jokipii}. Specifically, the pitch-angle diffusion coefficient  is prescribed following \citet{2013Agueda} and \citet{Wijsen2019(1)} as,
    
    \begin{equation}
        D_{\mu\mu} = D_{0}\left( \frac{|\mu|}{1+|\mu|}+\epsilon\right)(1-\mu^{2})
    \end{equation}
    
    where $\mu$ denotes the cosine of the pitch-angle and $\epsilon = 0.048$ prevents the occurrence of a resonance gap at $\mu = 0$. The coefficient $D_{0}$ is determined by relating $D_{\mu\mu}$ to the parallel mean free path $\lambda_\parallel$ through \citep{1970Hasselmann}
    
    \begin{equation}
    \lambda_\parallel = \frac{3v}{8}\int_{-1}^{1}\frac{\left( 1 - \mu^2\right)^2}{D_{\mu\mu}}d\mu.
    \end{equation}
    
    In this work, we assume that $\lambda_{\parallel}$ is constant in space and depends on the particle rigidity $R$ as
    \begin{equation}
        \lambda_{\parallel} = \lambda_{\parallel_{0}}  \left(\frac{R}{R_{0}}\right)^{(2-q)},
        \label{eq:MFP_//}
    \end{equation} 
    where $q=5/3$ is the Kolmogorov spectral index. The parameters $\lambda_{\parallel_{0}} = 0.3~\mathrm{au}$ and R$_{0}$ are the parallel mean free path and the rigidity for a particle with a given reference energy of 1~MeV.

    The perpendicular mean free path used for this study was chosen as:
    
    \begin{equation}
    \lambda_{\perp} = \frac{\alpha}{4 \pi} \frac{r_{g}}{r_{g_{\mathrm{0}}}}\lambda_{\parallel}.
    \label{eq:MFP_perp}
    \end{equation}
    
    Here, $r_{g}$ denotes the gyro-radius, $\alpha = 10^{-3}$ is a scaling factor, and $r_{g_{\mathrm{0}}}$ is the gyro-radius of a 1~MeV reference proton in a magnetic field of 5~nT, which is the value of the magnetic field at 1~au in the EUHFORIA background solar wind. A similar perpendicular mean free path was previously used by e.g., \citet{2010Dröge} and \citet{Wijsen2019(1)}.

\section{PARADISE complementary simulations}
\label{sec:complementary_PARADISE}

\begin{figure*}[htp]
        \centering
        \includegraphics[width=0.98\textwidth]{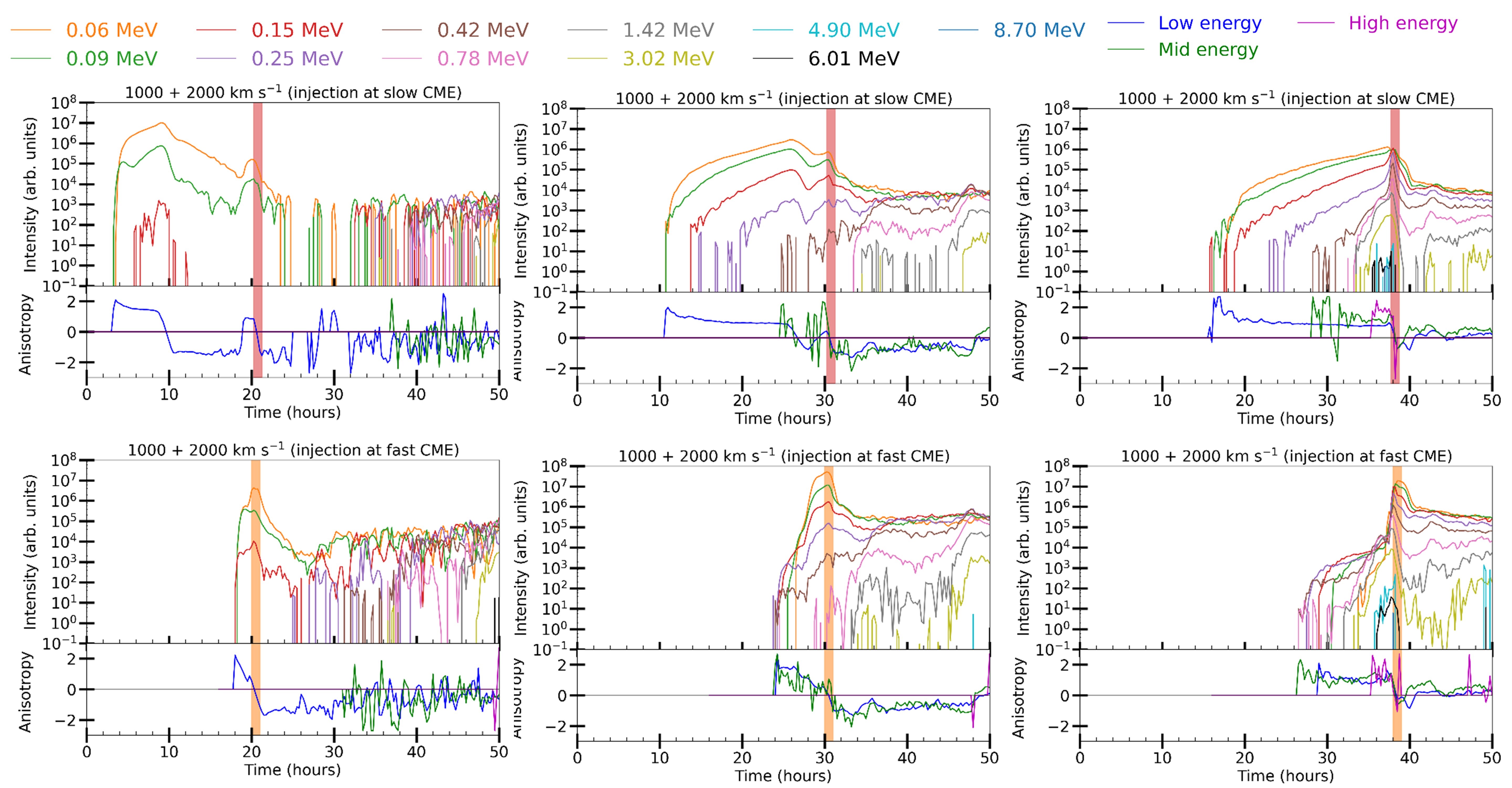}
        \caption{PARADISE complementary simulations results using the background solar wind with both CMEs injected. Panels~a (at 0.3~au),b (at 0.7~au), and c (at 1~au) show the resulting intensity and anisotropy time profiles when injecting only at the slower CME-driven shock position. Panels~d (at 0.3~au), e (at 0.7~au), and f (at 1~au) show the results when injecting only at the faster CME-driven shock position. Each panel consists of omnidirectional intensity-time profiles (top panel) and first-order parallel anisotropy time profiles (bottom panel). Low, mid, and high energy anisotropy-time profiles are calculated using the lowest 4, middle 4, and highest 4 energy channels, respectively. Vertical shaded areas correspond to a time window of $\pm$30~minutes centered at the time of switch in the low energy anisotropy sign, which, approximately, represents the time of the shock passage. 
        \label{fig:time_series_appendix}}
    \end{figure*}

    Figure~\ref{fig:time_series_appendix} shows the PARADISE simulation results when using EUHFORIA solar wind with both CMEs but injecting only in the slower CME (panels~a, b, and c) or injecting only on the faster CME (panels~d, e, and f). 

    When particles are injected in the slower CME, some of them propagate downstream of the CME-driven shock and may eventually reach the faster CME's shock. 
    This shock can further accelerate and mirror the particles, which explains the second peak in the registered intensities at 0.3~au in panel~a) of Fig.~\ref{fig:time_series_appendix}, approximately 20~hours after the injection of the slower CME. This is only visible in the lower energy channels because the slower CME is not as efficient at accelerating higher energy particles. This second peak is not as intense or energetic as the one seen in Fig.~\ref{fig:time_series} panel~g), as the particles are only injected at the slower CME.
     
    At 0.7~au (panel~b of Fig.~\ref{fig:time_series_appendix}) this second peak is registered approximately 30~hours after the injection of the slower CME and approximately 4~hours after the passage of the slower CME. As the particles that are propagated downstream of the slower CME-driven shock (and into the faster CME-driven shock) are trapped in between two shocks, they suffer from multiple acceleration processes, which translates into intensities not dropping into background levels as soon as the slower CME passes this position. 
    
    For the observer positioned at 1~au (panel~c of Fig.~\ref{fig:time_series_appendix}), very similar intensity-time profiles are registered for the first $\sim$35~hours compared to panel~i) of Fig.~\ref{fig:time_series}. This is a result of the continuous injection and acceleration. At this point in time, this observer is connected to a position where both shocks have already merged and higher energy particles are being produced.  Analogous to the process described for panel~i) of Fig.~\ref{fig:time_series}, as the faster shock approaches the slower shock, the MCMT process contributes to the acceleration of particles to higher energies. These particles serve as a seed population for the stronger merged shock, which is capable of accelerating them to even higher energies. This is reflected in the rapid rise of the high-energy anisotropy and the sharp rise in the intensities of these energies until the passage of the resulting shock.

    On the other hand, when particles are injected at the faster shock wave position, the profiles at 0.3 and 0.7~au (panels~d and e of Fig.~\ref{fig:time_series_appendix}) are very similar to the ones presented in panels~d and e of Fig.~\ref{fig:time_series}. In contrast to the previously described case, no second peak is registered at any of the different locations analyzed. Even though particles are not injected at the slower CME-driven shock, its impact on particle propagation is discernible for the observer at 0.7~au. Specifically, a minor intensity peak forms, centered on the slower CME's shock. The effect of the first CME is also clear at 1~au (Fig.~\ref{fig:time_series_appendix} panel~f), where the lower energy channels (0.05 and 0.09~MeVs) have their onset $\sim$2~hours later in Fig.~\ref{fig:time_series_appendix} panel~f) than in Fig.~\ref{fig:time_series} panel~f). As described for the top panels of Fig.~\ref{fig:time_series_appendix} once the shocks merge, the injection of particles only happens in the position of this new shock. Before the formation of the new shock, a fraction of the particles are trapped in between both CME-driven shocks (a pre-accelerated seed population), as the faster CME moves closer and closer to the position of the slower CME, these particles experience multiple acceleration processes, which result in higher energies being registered at 1~au. The maximum energy registered for this case corresponds to the 6.01~MeV energy channel, while when using the background with only the faster CME in the domain (Fig.~\ref{fig:time_series} panel~f), the maximum energy registered corresponds to the 3.02~MeV energy channel.

\end{document}